\documentclass[showpacs,preprintnumbers,amsmath,amssymb]{revtex4}

\usepackage{verbatim}

\usepackage{graphicx}
\usepackage{dcolumn}
\usepackage{bm}

\def\veps{\varepsilon}

\def\pa{\partial}

\def\calf{{\cal F}}
\def\calg{{\cal G}}

\def\call{{\cal L}}

\def\calo{{\cal O}}

\def \lc {\left(}     

\def \rc {\right)}     
\def \rC {\right]}

\def\beq{\begin{equation}}
\def\ee{\end{equation}}
\def\eeq{\end{equation}}
\def\pa{\partial}

\def\veps{\varepsilon}

\def\bfig{\begin{figure}}
\def\efig{\end{figure}}
\def\bea{\begin{eqnarray}}
\def\bwt{\begin{widetext}}
\def\ewt{\end{widetext}}
\def\beann{\begin{eqnarray*}}
\def\eea{\end{eqnarray}}
\def\eeann{\end{eqnarray*}}

\def\nn{\nonumber}

\def\3p0{$^{3}P_{\,0}$}

\begin{document}
 
\title{Astrophysical aspects of milli-charged dark matter in a Higgs-Stueckelberg 
model}
 
\author{A. L. dos Santos,  D. Hadjimichef}

\affiliation{Instituto de F\'{\i}sica, Universidade Federal do Rio Grande do
  Sul, Av. Bento Gon\c{c}alves, 9500\\
Porto Alegre, Rio Grande do Sul, CEP 91501-970,
Brazil
} 

\begin{abstract}
An extension of the Standard Model is studied, in which two new vector bosons are introduced, a first boson $Z'$ coupled to the SM by the usual minimal coupling, producing an enlarged
gauge sector in the SM. The second boson $A'$ field, in the dark sector of the model, remains massless and originates a dark
photon $\gamma'$. A hybrid mixing scenario is considered based on a combined Higgs and Stueckelberg
mechanisms.  In a Compton-like process a photon scattered by a WIMP
is converted into a dark photon. This process is studied, in an astrophysical application obtaining
an estimate of the impact on stellar cooling of white dwarfs and neutron stars.
\end{abstract} 

\pacs{95.35.+d, 12.60.-i}
\maketitle

\section{Introduction}
\label{intro}
 
The enigma of dark matter (DM) remains unsolved.
Over the years, dark matter candidates converged to a
variety of interesting and plausible candidates namely
the {\it weakly-interacting massive particles} (WIMPs) \cite{silk}. In
general they are present in theories of weak-scale physics
beyond the Standard Model (SM) and give rise to appropriate relic abundance. Calculations have shown that
stable WIMPs can remain from the earliest moments of
the Universe in sufficient number to account for a significant fraction of relic dark matter density. This raises
the hope of detecting relic WIMPs directly by observing
their elastic scattering on targets. In the dark matter zoo
many different types of particles have been introduced
and their properties theoretically studied.

Still the most promising candidates are supersymmetric dark matter particles \cite{jungman}, although other non-baryonic
candidates have been proposed \cite{dodelson}-\cite{dm1}. If a fermionic dark
matter candidate is subject to standard Fermi interactions, Lee and Weinberg have shown that relic density
arguments precludes a WIMP with a mass lower than a
few GeV \cite{lee}. However, it has been suggested \cite{integral}, related
to the INTEGRAL satellite observations of the galactic
bulge, that the gamma-ray emission line of 511 keV could
be the product of light dark matter particles annihilating into positrons which then annihilate producing the
observed gamma-ray radiation.

Alternatively, many types of models that explore the
physics beyond SM, share in common the presence of new
$U(1)$ vector bosons \cite{langacker}. These new bosons are introduced
basically in two ways: (i) minimal coupling; (ii) Stueckelberg mechanism. A new vector gauge boson would be
massless if a new $U(1)$ symmetry should remain unbroken. This would imply in a long range force if it were
to couple to ordinary matter, unless the coupling were
incredibly small. As shown by Dobrescu \cite{dobrescu},\cite{dobrescu2} this case
would be allowed if the primary coupling were to a hidden
sector and connected only by higher-dimensional operators or alternatively by kinetic mixing with the photon.
In the case of kinetic mixing, this scenario would induce a
small fractional electric charge for hidden sector particles
\cite{holdom}. 
Additionally, any new $U (1)$ gauge symmetry  must be anomaly free. 
An important issue is how these anomalies are ultimately canceled. The two main anomaly-cancellation 
scenarios then divide according to whether or not anomalies cancel among the
SM fields themselves, or require the addition of new particles. These cases have been studied in the
literature \cite{anomal1}-\cite{nath1}. 
Numerous models describing possible physics beyond the SM predict the existence of narrow resonances
at′ the TeV mass scale. Results of searches for narrow
$Z \rightarrow l^+ l^-$ in pp collision data have previously been reported by the ATLAS \cite{atlas} and CMS \cite{cms} collaborations.

In our present study we define a Higgs-Stueckelberg
extension of the SM where in the dark sector one has
a QED-like model 
which consists of a fermion singlet ($\chi$) and a
dark photon ($\gamma'$). The connection between the dark sector 
and the  SM is  accomplished by an extra heavy gauge
boson ($Z'$). The $Z'$ couples to the SM by minimal coupling and to
the dark QED-like sector
by the known Stueckelberg mechanism. This coupling 
renders some interesting features
such as the SM particles are neutral to $\gamma'$ . If a dark
QED could exist, independently, with a similar physical
and mathematical structure as ordinary QED, what effects could it produce on SM particles? 
The presence of an extra $Z'$ boson is crucial to establish a link between this new QED-like model and the
original one incorporated in the Standard Model. There
are many test grounds that could be explored, we shall
investigate a simple energy loss mechanism in stars. In
this direction, similar calculations have been performed
in dense stellar matter, for white dwarfs and central protoneutron star (PNS)  in supernova scenarios.

In white dwarfs, an extra cooling mechanism has been taken into account 
considering a very light dark matter candidate, the axion, originally introduced as an attempt to solve the CP violation
problem \cite{cp}. In supernovas,  neutrino radiation, on time scales of tens of seconds, during
which the central protoneutron star (PNS) cools deleptonizes, and contracts. 
Our calculation introduces an alternative
mechanism for the energy loss which can be tested, in model building grounds and, in principle, could
work together with the axion physics for white dwarfs and neutrino cooling in neutron stars.

\section{The model}
\label{sec:dQED}

There is now a large body of evidence for the existence of  dark matter that
interacts gravitationally and makes up nearly a quarter of the energy density of the universe.
Although little is known about it, a possible framework for studying its properties is in a simplified model.
 Many models have been proposed, sharing a similar structure: a Standard Model (SM) sector, a dark matter sector and the interaction between these sectors.  In general, this interaction is called a {\it portal}, which is communication link among the particle sectors. In the literature, 
there are some extensively studied examples: Higgs portal \cite{portal1}-\cite{portal5}, Fermion portal \cite{portal6},  
here we examine another possibility: a $U(1)$ vector boson  portal. 
The dark sector of our model is extremely simple, composed of a fermion $\chi$ and a vector boson $X_\mu$. Gauge invariance would require an extra
Higgs field for mass terms, which would introduce new unknown parameters.   A alternative solution for mass terms,  vector bosons and gauge invariance is  the Stueckelberg mechanism, where an unphysical field $\sigma$ is introduced, but decoupled after gauge fixing.
In this sense, the new boson $C_\mu$ can be defined in a  hybrid coupling scenario: first  $C_\mu$   couples to the SM by the
usual minimal coupling, producing an enlarged gauge sector in the SM  (enSM),
 then in the dark sector,   $C_\mu$  mixes $X_\mu$   via Stueckelberg coupling. 
 
\begin{figure}[t]
\centering
\includegraphics[angle=-90,width=.9\columnwidth ]{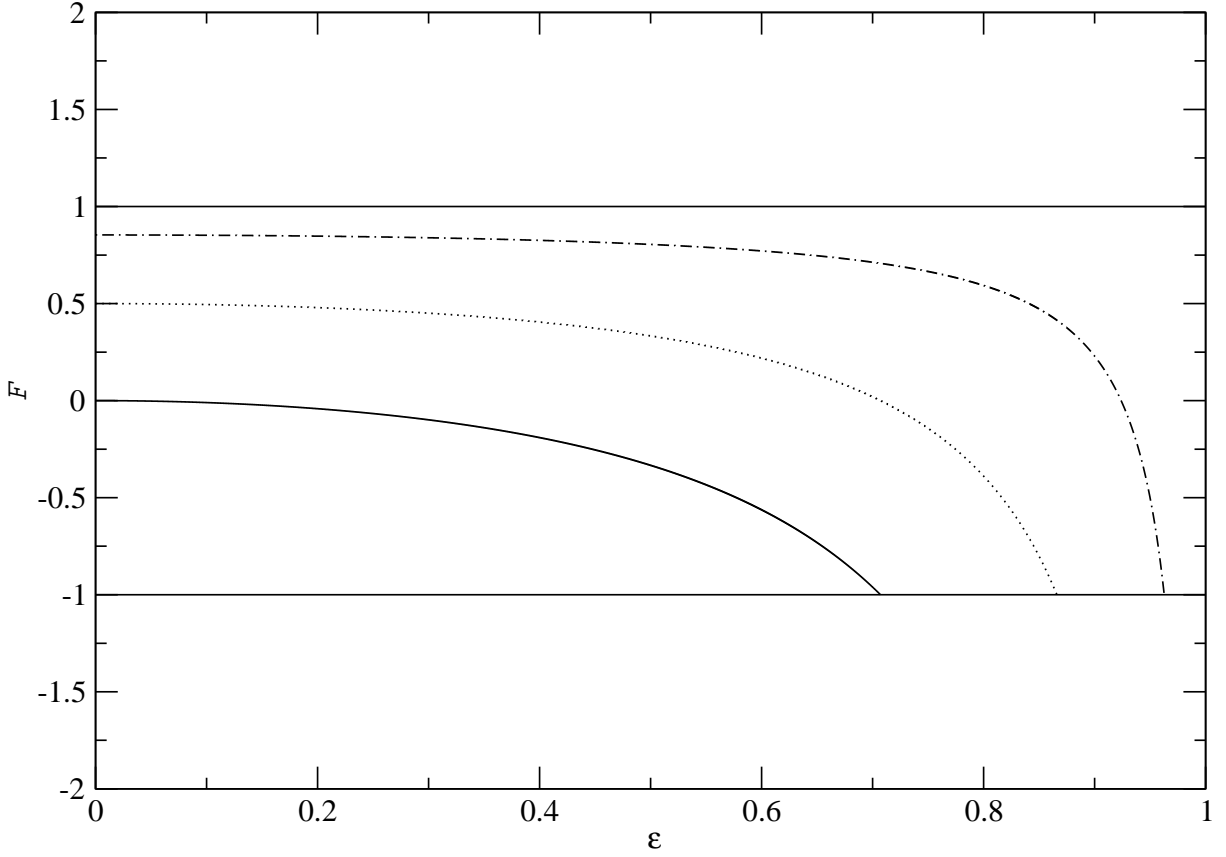}
  \caption{The allowed values of $m_{Z'}$ for $\phi=\pi/8$ (dash-dot), $\phi=\pi/4$ (dash) and $\phi=\pi/2$
(solid). }
\label{f}
\end{figure}

\begin{figure}[t]
\centering
\includegraphics[width=.9\columnwidth]{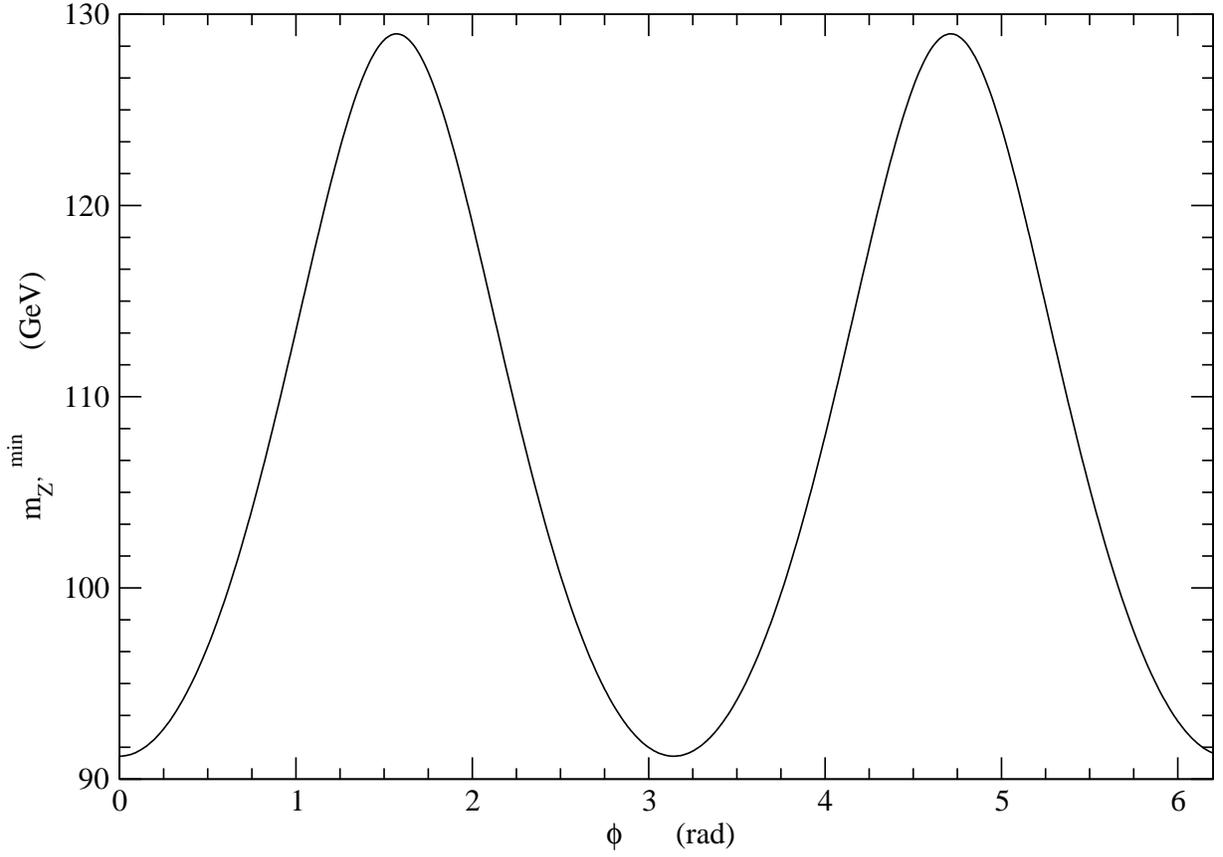}
  \caption{The minimum values for $m_{Z'}$.  }
\label{min}
\end{figure}

The extra symmetry groups are defined by ${\cal G} = U (1)_C \,\otimes U (1)_X$ , so the model Lagrangian density
can be written as
\bea
\call = \call_{\rm \,enSM}  +    \call_{\rm St} +\call_\chi    \,.
\label{lag}
\eea
The   model is defined by
\begin{subequations}
\bea
\call_{\rm \,enSM}&=&-\frac{1}{4}\,W^a_{\mu\nu}\,W^{a \mu\nu}-\frac{1}{4}B_{\mu\nu}\,B^{ \mu\nu}-\frac{1}{4}C_{\mu\nu}C^{ \mu\nu}
+(D_\mu\Phi)^\dag\,D^\mu\,\Phi-V(\Phi^\dag\Phi)
+ i\bar{\psi}_f\, \gamma^\mu D_{\mu} \, \psi_f\,
\label{dm0}
\\
\call_{\rm St}&=&-\frac{1}{4}X_{\mu\nu}X^{ \mu\nu}
+\frac{1}{2}\,\Lambda_\mu\,\Lambda^\mu
\label{dm1}
\\
\call_\chi&=& i\bar{\chi}\left( \gamma^\mu \pa_{\mu} -m_\chi\right) \chi\,+\call^\chi_{\rm int}\,,
\label{dm1b}
\eea
\end{subequations}
with
\begin{subequations}
\bea
-\call^\chi_{\rm int}&=& g_x\,Q_x\,X_\mu\,j^\mu_\chi
\label{dm2a}\\
D_{\mu}&=&
\partial_{\mu} + i\, g_2 \,\frac{\tau^{\,3}}{2}   W^{3}_{\mu}
+i\, g_{Y}\,\frac{Y}{2}\, B_{\mu}
+i\, g_{C} \frac{Y}{2}  \, C_{\mu}\,,
\label{dm2}
\eea
\end{subequations}
where $j^\mu_\chi=\bar{\chi}\,\gamma^\mu\,\chi$ and in $D_{\mu}$ only the relevant part, that couples the vector bosons, is presented.
The new $U(1)$ bosons define the field tensors $X_{\mu\nu}=\pa_\mu\,X_\nu-\pa_\nu\,X_\mu$ and  
$C_{\mu\nu}=\pa_\mu\,C_\nu-\pa_\nu\,C_\mu$. The second term
in (\ref{dm1}) is the Stueckelberg mixing term between the 
two  boson fields $C_\mu$ and $X_\mu$ via an axial pseudo-scalar $\sigma$ field given by
\bea
\Lambda_\mu&=&\partial_\mu \sigma -m_1\, C_\mu - m_2 \,X_\mu\,.
\label{dm3}
\eea
The  term in (\ref{dm1b})  is a  fermion singlet term of the dark sector. 
The $\sigma$ field  is   unphysical and decouples from all fields after gauge fixing
\bea
\call_{\rm gf}=-\frac{1}{2\xi}\left(\pa_\mu C^\mu+\xi\, m_1\, \sigma\right)^2
-\frac{1}{2\xi}\left(\pa_\mu X^\mu+\xi\, m_2\, \sigma\right)^2\,.
\label{lgf}
\eea
This type of  model was first proposed by \cite{nath1}, but with just one dark field, 
and was applied as well in \cite{taiwan}. 
Using this full Lagrangian (\ref{lag}) and taking into account just the terms that contribute 
to the vector bosons masses, we write the relevant term that corresponds to 
the squared mass matrix 
\bea 
\call_{\rm M}= +\frac{1}{2}
V^{\mu\,T}\,
 M^2\, V_\mu
\,,
\eea 
where $V^{\mu\,T}=\left(X^\mu, C^\mu, B^\mu, W^{3 \mu} \right)$, the squared mass matrix becomes
\bea 
M^2=
\lc 
\begin{array}{cccc}
 m_2^2 & m_1 m_2 & 0 & 0 \\
 m_1 m_2 & m_1^2+\frac{g_{C}^2 v^2}{4} & \frac{1}{4} g_{C} g_Y v^2 & -\frac{1}{4} g_2
   g_{C} v^2 \\
 0 & \frac{1}{4} g_{C} g_Y v^2 & \frac{g_Y^2 v^2}{4} & -\frac{1}{4} g_2 g_Y v^2 \\
 0 & -\frac{1}{4} g_2 g_{C} v^2 & -\frac{1}{4} g_2 g_Y v^2 & \frac{g_2^2 v^2}{4}
\end{array}
\rc.
\nn\\
\label{mass}
\eea 
It is convenient to define
\bea
 m_W &=& \frac{g_{2}\,v}{2}
\,\,\,\,\,\,;\,\,\,\,\,\,
m_C = \frac{ g_{C}\,v}{2}
\,\,\,\,\,\,;\,\,\,\,\,\,
m_Y = \frac{ g_Y\,v}{2}\,;
\nn\\
m_Z^2&=&m_W^2+m_Y^2
\,\,\,\,\,\,;\,\,\,\,\,\,
m_{Z'}^2=m_1^2+m_2^2
\label{defs}
\eea
and introduce the following parametrization
\bea
\tan{\phi}&=&\frac{m_1}{m_2}\,. 
\label{tan}
\eea
After diagonalizing the matrix (\ref{mass}) we obtain four   mass-squared eigenvalues
$M_1^2$, $M_2^2$, $M^{2}_{+}$ and $M^{2}_{-}$:
\begin{subequations}
\bea
m_\gamma^2  &\equiv& M_1^2  = 0
\,\,\,\,\,\,\,\,;
\,\,\,\,\,\,\,\,
m_{\gamma'}^2   \equiv M_2^2 = 0
\label{mass12}\\
M^2_\pm&=&\frac{1}{2} \left[ m_Z^2 + m_{Z'}^2+m_{C}^2 \right]   \pm\,\frac{1}{2}\Delta 
\label{mass1}
\eea
where
\bea
\Delta=\sqrt{
m_C^4  + 2 m_C^2\, ( m_Z^2 -  m_{Z'}^2\cos{2 \phi})+ (m_Z^2 - m_{Z'}^2)^2
}
\nn
\eea
\end{subequations}
The experimental value for the mass of Z boson is \cite{PDG}
\bea
\overline{m}_Z = m_Z \pm \delta\,m_Z = 91.1876 \pm 0.0021\,\,\, \rm{(GeV).}  
\eea
The above analysis requires that the effects of the
extended model on the $Z$ mass must be such that it should lie in, what was called, ``the error
corridor"  of the SM prediction \cite{nath1,taiwan,zp}. To calculate the error $\delta m_Z$ in the SM prediction of 
$m_Z$, we consider that the coupling $g_C$ is small  in order that $\Delta$ can be expanded in powers
of $m_C$. Expanding (\ref{mass1}) up to second order in $m_C$, one
finds
\begin{subequations}
\bea
M_{+}&\approx& m_{Z'}+ \frac{m_C^2}{2m_{Z'}}\,\,\calg(\phi\,,\veps) 
\label{mass41}\\
M_{-}
&\approx&
m_Z+ \frac{m_{C}^2}{2\,\,m_{Z}}\,\, \calf(\phi\,,\veps)\,,
\label{mass4}
\eea
\end{subequations}
where
\bea
\calg(\phi\,,\veps)&=&\frac{\sin^2\phi}{1-\veps^2}\,
\nn\\
\calf(\phi\,,\veps)&=&  \frac{\cos^2\phi-  \veps^2     }{  1- \veps^2  } 
\label{mass4a}
\eea
with $\veps=m_Z/m_{Z'}$\,. In the simple case $\phi=0$, Stueckelberg decouples from the enSM and $\calf(0,\veps)=1$.
To lie in the error corridor of the SM prediction, 
$m_C\approx 603.9$ MeV, which implies that $g_C$, with $v=246$ GeV, has maximum value of
\bea
g_C=\frac{2\,m_C}{v}\approx 5\times 10^{-3}\,.
\label{mass7}
\eea
 Coupling  Stueckelberg to the enSM implies  that $\phi\not=0$. In this case the effective $g_C$ 
differs from the value obtained in (\ref{mass7}) by the factor
 $\calf$ present in Eq. (\ref{mass4}) restricted to span the range
\bea
 -1 \leq \calf(\phi\,,\veps) \leq 1\,. 
\label{mass8}
\eea
In figure (\ref{f}) the mass curves for $m_{Z'}$, for three angles ($\phi=\pi/8,\pi/4,\pi/2$), are plotted.
As can be seen, there is a cut-off value for each chosen angle, which corresponds to a minimum value for
an experimental search of $m_{Z'}$. 
The model can not determine the ${Z'}$ mass, but a lower bound value can be obtained. 
This  value can be calculated exactly, solving the equation 
$\calf(\phi\,,\veps)=-1$ for $\veps$, obtaining
\bea
m_{Z'}^{\rm  min}=m_Z\,\,\sqrt{\frac{2}{\cos^2\phi+1}  }\,,
\label{mass10}
\eea
which is plotted in figure  (\ref{min}). The minimum value for $m_{Z'}$, in this model, oscillates as a function of $\phi$.
In a  consistent $Z'$ hunt one  first  fixes the angle $\phi$, using Eq. (\ref{mass10}), at a definite value,
obtaining the corresponding lower bound  $m_{Z'}^{\rm  min}$. This  implies that the true $Z'$ mass can not be lighter than this value.
 For example, if $\phi=0$ the minimum value for the $Z'$ mass is simply   $m_Z$, which suggests that no 
$Z'$ lighter than the usual $Z$ boson is allowed.
Now for  $\phi=\pi/2$  (largest minimum) one has $m_{Z'}^{\rm  min}=128.97$ GeV and, for this angle, no 
lighter $Z'$ mass value is possible.

A $4\times 4$ orthogonal matrix   $\calo$
can be obtained such that it diagonalizes  (\ref{mass})   
and transforms the basis, connecting physical 
fields $F_\mu=(A^{\prime}_\mu, Z'_\mu,Z_\mu, A_\mu)$ 
with  $V_\mu$
\bea
V_\mu=\calo\,F_\mu\,.
\label{o-fields}
\eea 
Considering parametrization (\ref{tan}) and the  following 
\bea
\tan{\eta}&=&\frac{g_{C}}{g_Y}\cos{\phi}
\nn\\
\tan{\theta}&=&\frac{g_Y^{\prime}}{g_2}=\frac{g_Y}{g_2} \sec{\eta}
\nn\\
\tan{\psi} &=& \frac{ m_W^2\sin{\eta} \tan {\phi } \sin {\theta}}{m_W^2-m^2_{Z'} \cos^2{\theta} }
\label{angles}
\eea
with $g_Y^{\prime}=\sqrt{g_Y^2+ g_C^2 \cos{\phi}^2}$, the
$\calo$ matrix    can be determined in terms of these quantities, such that  
%
%
%
%
\bea 
\calo=
\lc 
\begin{array}{cccc}
 c_\eta s_\phi\,\, 
& c_\phi c_\psi+s_\eta s_\phi s_\psi s_\theta \,\,\, 
& c_\phi s_\psi-c_\psi s_\eta s_\phi s_\theta \,\,\,
&-c_\theta s_\eta s_\phi     
\\
-c_\eta c_\phi 
& c_\psi s_\phi-c_\phi s_\eta s_\psi s_\theta 
& s_\phi s_\psi+c_\phi c_\psi s_\eta s_\theta 
& c_\phi c_\theta s_\eta 
\\
s_\eta 
& -c_\eta s_\psi s_\theta       
& c_\eta c_\psi s_\theta     
&    c_\eta c_\theta   
\\
0     
&    
c_\theta s_\psi    
& -c_\psi c_\theta     
& s_\theta
\end{array}
\rc\,,
\nn\\
\label{mass-param}
\eea 
with $c_\theta = \cos{\theta}$, $s_\theta = \sin{\theta}$ etc. The transformed fields are written as
\bea
X_\mu &=& c_\eta s_\phi A'_\mu + (c_\phi c_\psi+s_\eta s_\phi s_\psi s_\theta) Z'_\mu 
+ (c_\phi s_\psi-c_\psi s_\eta s_\phi s_\theta) Z_\mu -c_\theta s_\eta s_\phi A_\mu 
\nonumber\\
C_\mu &=& -c_\eta c_\phi A'_\mu + (c_\psi s_\phi-c_\phi s_\eta s_\psi s_\theta) Z'_\mu 
+  (s_\phi s_\psi+c_\phi c_\psi s_\eta s_\theta) Z_\mu + c_\phi c_\theta s_\eta A_\mu 
\nonumber\\
B_\mu &=& s_\eta A'_\mu -c_\eta s_\psi s_\theta Z'_\mu +  c_\eta c_\psi s_\theta Z_\mu + c_\eta
   c_\theta A_\mu 
\nonumber\\
W_{\mu}^3 &=& c_\theta s_\psi Z'_\mu -c_\psi c_\theta Z_\mu + s_\theta A_\mu .
\label{camps}
\eea
The derivative defined in (\ref{dm2}) can be written using (\ref{camps})
\bea
D_{\mu}=\partial_{\mu} +  
\frac{i}{2}\,\left(\alpha_1\,A'_\mu+ \alpha_2\,A_\mu+ \alpha_3\,Z'_\mu+\alpha_4\,Z_\mu \right)
\label{derivada1}
\eea
where 
\bea
\alpha_1&=& 0
\nn\\
\alpha_2&=& 
e\, Q_{em}
\nn\\
\alpha_3&=& g_2\,s_\psi \,c_\theta\,\tau^{\,3}  
-\left[\,\frac{}{}
g_Y\,c_\eta\,s_\psi\, s_\theta\,
-g_{C}\,\left( s_\phi\,c_\psi -   c_\phi\,s_\psi\,s_\eta\,s_\theta\,\right) 
\right]\,Y
\nn\\
\alpha_4&=& -g_2\,c_\psi\, c_\theta\,\tau^{\,3}  +  
\left[\,\frac{}{} g_Y\,c_\eta\,c_\psi\, s_\theta\,
+g_{C}\,\left( s_\phi\,s_\psi +   c_\phi\,c_\psi\,s_\eta\,s_\theta\,\right)
\right]\,Y\,
\label{alphas}
\eea
with $Q_{em}\,=\,\tau^{\,3}/2 +Y/2$ is the usual charge operator and
the eletric charge is defined as
\bea
e&=&\frac{g_Y^{\prime}g_2}{ \sqrt{g_2^2+(g_Y^\prime)^2  } }\,.
\label{cargas2}
\eea
As a direct consequence of $\alpha_1=0$ in (\ref{alphas}), 
the interaction of a dark photon field $A^{'}_{\mu}$ with a SM model current $j_\mu$
has no contribution and guarantees  no direct coupling between  the dark photon  and SM charged particles.
The interaction Lagrangian in the dark sector, defined in (\ref{dm2a}), can be written using (\ref{camps}) as
\bea
-\call^\chi_{\rm int}&=& g_x\,\left(\alpha_5\,A'_\mu+ \alpha_6\,A_\mu+ \alpha_7\,Z'_\mu+\alpha_8\,Z_\mu \right)\,j^\mu_\chi
\label{derivada2}
\eea
where
\bea
\alpha_5&=& g_x\,c_\eta s_\phi\,Q_x
\nn\\
\alpha_6&=& -g_x\,c_\theta s_\eta s_\phi \,Q_x
\nn\\
\alpha_7&=&  g_x\,(c_\phi c_\psi+s_\eta s_\phi s_\psi s_\theta)  \,Q_x
\nn\\
\alpha_8&=&  g_x\,(c_\phi s_\psi-c_\psi s_\eta s_\phi s_\theta)     \,Q_x\,.
\eea
The coupling of the photon to the $\chi$ fermion of the dark sector results in a dark charge $e_x$
\bea
e_x=g_x  c_\eta s_\phi
=\xi\, e
\label{ex1}
\eea
where 
\bea
\xi = \frac{g_x g_Y}{g_2}\sin{\phi} \frac{\sqrt{g_2^2+g_Y^2+g_C^2 \cos^2\phi}}{g_Y^2+g_C^2 \cos ^2\phi}.
\label{xi1}
\eea
To  guarantee dark charge conservation, the dark photon  has the same coupling to the $\chi$ fermion as the ordinary
photon. The dark charge is dependent on the degree of mixture of the SM with the dark sector. In the case of $\phi=\pi/2$ one has
$g_{Y'}=g_{Y}$, using the known values $g_2\simeq 0.65$   and $g_Y\simeq 0.35$,
 Eq. (\ref{xi1}) simplifies to
\bea
\xi =g_x \,  \frac{\sqrt{g_2^2+g_Y^2}}{g_Y \, g_{2}}\,\,\simeq\,\,3.2\,\,g_x.
\label{xi2}
\eea
If the coupling $g_x$ is small, for example, the same order of $g_C$, then the dark electric charge is a fractional
{\it milli-charge}, $e_x\sim 10^{-3}\, e$\,. Historically, the context of milli-charged dark matter was first 
discussed by Holdom \cite{holdom}, Goldberg and Hall \cite{goldberg} and in recent studies \cite{taiwan}-\cite{zp}.
In summary, these studies have shown that milli-charged particles, with  fractional electric charge ranging from
$10^{-6}$ to $10^{-1}$  of a unit charge are  allowed.


\section{An astrophysical application}

\subsection{Stellar energy loss}

\begin{figure}
\centering
\includegraphics[width=1.0\columnwidth]{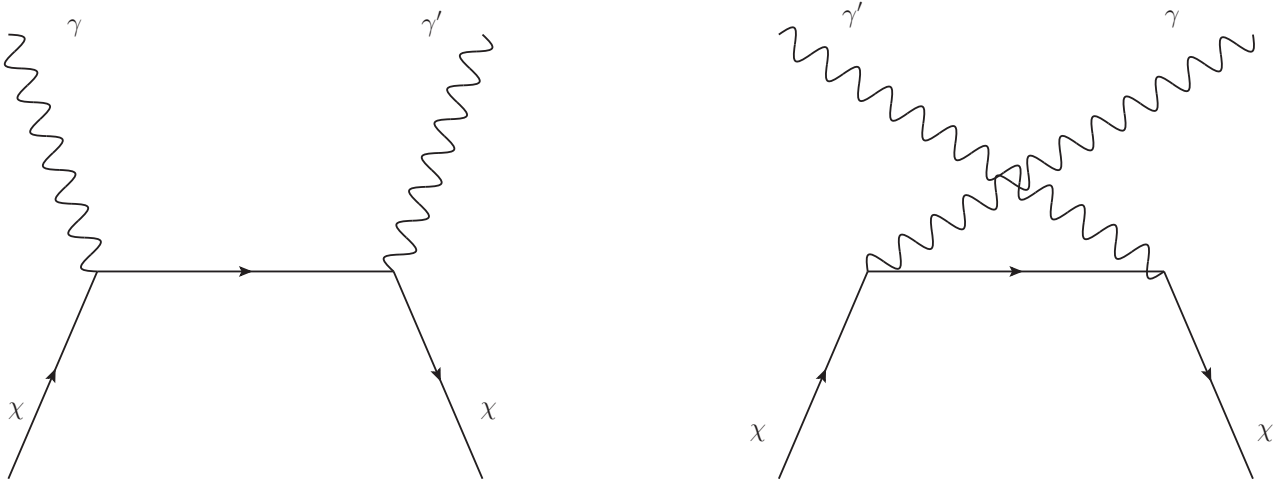}
  \caption{Compton diagrams}
\label{compton-diag}
\end{figure}

Astrophysical observations have become a well-known tool to obtain empirical constraints 
on new particles.
Any light particle,   in principle, has a potential for playing an important
role in stellar energy loss. Such a particle would remove energy from the  stellar thermal
bath by a direct mechanism. 
If stellar
matter has a sufficient content of dark matter, an important process to consider, is the $\gamma'$
emission from thermal states. This is relevant for determining the relic cosmological abundance, but is also the
source of important constraints arising from new energy-loss mechanisms in stars. To obtain an estimate of the
impact on stellar cooling, we will focus on the Compton-like process
$\chi+\gamma\to\chi+\gamma^\prime$  by the diagrams in figure (\ref{compton-diag}). The
energy loss is given by $Q_{\gamma'}$
\bea
Q_{\gamma'} = \frac{1}{\rho} \int \frac{d^3 q_\gamma}{(2 \pi)^3} \frac{2}{e^{\omega \beta} -1} 
\int \frac{d^3 q_{\chi}}{(2 \pi)^3} \frac{2}{e^{E_{\chi} \beta} +1} \,\,\,\sigma_c \,E_{f}\,\,
\label{qgp}
\eea
where $\omega$ is photon energy in the thermal bath; $E_{\chi}$ is the energy of the fermions of the dark sector; 
$E_f$  is the star's  energy loss due to the dark photons and $\sigma_c$ is the Compton cross section of the processes. 
The total cross-section for dark photon production  from  Compton scattering is 
\bea
\sigma_c &=&  \pi  \frac{(\xi^2\alpha)^2}{ m_\chi}  
\Bigg[  \frac{4 m_\chi}{\omega^2} + \left.
2\frac{(m_\chi+\omega)}{( m_\chi+2 \omega)^2}
-\frac{1}{\omega^3}( 2 m_\chi^2+2 m_\chi \omega -\omega^2) \ln\lc{1+\frac{2 \omega}{m_\chi}}\rc\rC
.
\label{compton}
\eea
To evaluate    $Q_{\gamma'}$ in (\ref{qgp}), one can use  the following:  $E_f \approx \omega$, 
$d^3 q_\gamma = dk |\vec{k}|^2 d \Omega$, 
$d^3  q_{\chi} = dp |\vec{p}|^2 d \Omega$, 
$dk = d\omega$, $dp = (E_\chi/|\vec{p}|)dE_\chi$, 
$\omega^2 = |\vec{k}|^2 $ and $|\vec{p}|^2 = E_\chi^2 - m_\chi^2 $.
The energy loss becomes
\bea
Q_{\gamma'} &=& \frac{1}{\rho} \frac{1}{(2 \pi)^6} \int_{0}^{\infty} (8\pi) d \omega\frac{\omega^3}{e^{\omega \beta} -1} 
\int_{m_\chi}^{\infty} (8\pi)d E_\chi \sqrt{E_\chi^2 - m_\chi^2}\frac{E_\chi}{e^{E_\chi \beta} +1} \,\sigma_c \,.
\eea
We approximate $\sigma_c$ as a constant when compared to the
integral of $E_\chi$, which results in
\bea
Q_{\gamma'}= \frac{ m_{\chi}^{5} (\xi^2\alpha)^2}{\rho\pi^3} \,I_1\,I_2
\label{qfoton}
\eea
where
\bea
I_1&=&\int_{0}^{\infty} \frac{dx}{e^{\beta\,m_\chi\,x} -1}   \,f(x)
\nn\\
I_2 &=&\int_{1}^{\infty} d x \,\sqrt{x^2 - 1}\,\frac{x}{e^{\beta\,m_\chi\,x} +1}
\eea
with
\bea
f(x)=  4x
+  \frac{2x^3( 1 + x  ) }{( 1 + 2 x  )^2}
 - (2  +2 x -x^2) \ln(1+2x)\,.
\nn
\eea
In order to establish a comparison of our results with other stellar cooling mechanisms 
we shall consider two known cases: axions in  white dwarfs   and neutrinos in  protoneutron stars.

The axion was originally introduced as 
a very light dark matter candidate,  
where the supernova 1987A dynamics and laboratory
searches has constrained its mass to values $\lesssim 0.01$ eV.
The main processes for axion emission are the Compton-type, the Primakoff process and the annihilation process
\cite{axion0}-\cite{axion3}. The comparison with the present calculation
(\ref{qfoton}) is established considering an axion Compton process $e + \gamma \rightarrow e + a$, extracted from \cite{axion0}

\bea
Q_a=5.29\times 10^{4}\frac{1}{\mu_e}\,T_8^{\,6}\,I(T_8,\rho)\,\left(\frac{m_a}{{\rm eV}}\right)^2\,,
\label{eq_axion0}
\eea
with (\ref{eq_axion0}) in ergs/g sec; $T_8$ the temperature in units of
$10^8$ K; $m_a$ the axion’s mass. The factor $I$ is tabulated
in Appendix A of \cite{axion0}, where $I = 1$ is the nonrelativistic
and nondegenerate limit. 
As discussed in \cite{axion0}, if an axion with mass $m_a$ of 1 eV
exists, the nuclear energy generation should be more than
a hundred times larger and then the time scale of evolution becomes shorter by that factor. This effect should
disturb the star distribution in the horizontal branch,
therefore this mass value can be considered as an upper
bound. By this fact, the value of $1$ eV is used as one of
the reference masses in the axion energy loss calculation.

The characteristic signatures of dark matter are potentially detectable with the analysis of the stellar oscillations. 
Asteroseismology is presently showing its power in
determining with high precision not only the global properties of stars but also their internal structure. Recently,
A. H. C\'orsico  et al. used the {\it state-of-the-art} asteroseimological model \cite{corsico1} to study the rate of the anomalous
cooling of the pulsating white dwarf star G117-B15A.
From this measure they inferred the axion mass due to
this extra cooling mechanism and obtained the value of $m_a \cos^2 \beta = 17.4_{-2.7}^{+ 2.3}$ meV, where $\cos^2 \beta$ is a free, 
model-dependent parameter that is usually set equal to unity
\cite{corsico2}. 

The energy loss  is density dependent quantity, therefore a $Q$ calculation  involves   an estimate
of   ordinary stellar matter density,  which is very well established 
for axions or neutrinos, both coupling to SM fields. 
Due to the restrictions imposed by  Eq. (\ref{alphas}) ($\alpha_1=0$)
on direct interactions of  dark photons with ordinary
SM charged particles,  a $\gamma\leftrightarrow\gamma'$ conversion must always involve a dark $\chi$ fermion. 
A $Q$   calculation in this scenario implies in a rough estimate of the dark matter  density.
On distance scales of the size of galaxies and clusters of
galaxies, evidence of dark matter are compelling, but still observations do
not allow one to determine the total amount of dark matter in the Universe.
Information has been extracted from the analysis of the cosmic microwave background (CMB).
In particular, stringent constraints on the abundances of baryons and matter in the Universe has been
placed by the Wilkinson microwave anisotropy probe (WMAP) data
and recently by high resolution detections of both the total intensity and polarization of primordial CMB anisotropies
by Planck \cite{planck}. $N$-body simulations suggest the existence of a universal dark 
matter profile for halo densities, where some of the most
widely used profile models are Kravtsov  et al.  \cite{profile1}, Navarro, Frenk and White  \cite{profile2}, Moore  et al. 
 \cite{profile3}  and modified isothermal  \cite{profile4} profiles.  
Therefore, if  dark matter can be assigned as  a constituent of stellar  interiors, in a first approximation, 
 a conservative choice for the unknown stellar density  is to  assume that it  matches 
the density of ordinary stellar matter. This is consistent with recent studies where ordinary stellar matter is mixed 
with   non-self-annihilating  dark matter \cite{mix1,mix2}. It was found that a more  compact  star results when a DM core is 
included. Dark matter density profiles are presented showing a high density dark matter stellar core. 
As will be shown, this is sufficient 
for the dark photon mechanism to be comparable with other known cooling  mechanisms.

\subsection{Astrophysical constraints}

Stellar density,
assumed to be made of pure hydrogen, ranges from $10^2 \lesssim \rho \lesssim 10^4$ g$\cdot$cm$^{-3}$ 
which is,  typically the range for a star like the   Sun  to a red giant. 
Compact stars like white dwarfs, have  higher densities of matter, of order
$10^6$ g$\cdot$cm$^{-3}$ and temperatures of $10^7$ K.
In the temperature range around $10^8$ K, the Compton-type
process (\ref{eq_axion0}) is dominant \cite{axion0}-\cite{axion3}.

\begin{figure}[t]
\centering
\includegraphics[angle=-90,width=.9\columnwidth ]{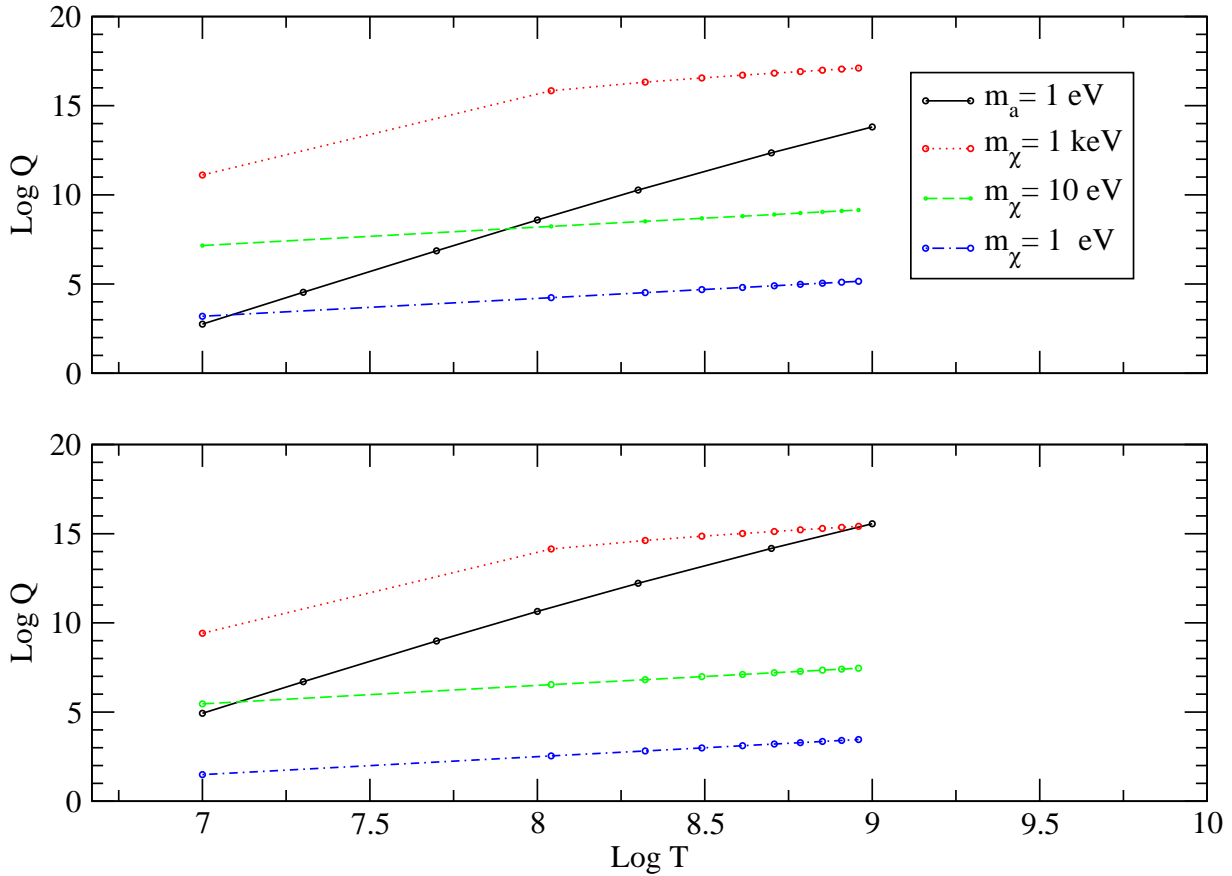}
  \caption{Comparison of the energy loss for an axion of mass $m_a$=1 eV and a fermion singlet $\chi$\,: $\rho\sim 10^4$ g/cm$^3$ (upper),
 $\rho\sim 10^6$ g/cm$^3$ (lower). Q in ergs/g.sec and T in K. }
\label{el}
\end{figure}

\begin{figure}[t]
\centering
\includegraphics[angle=-90,width=.9\columnwidth ]{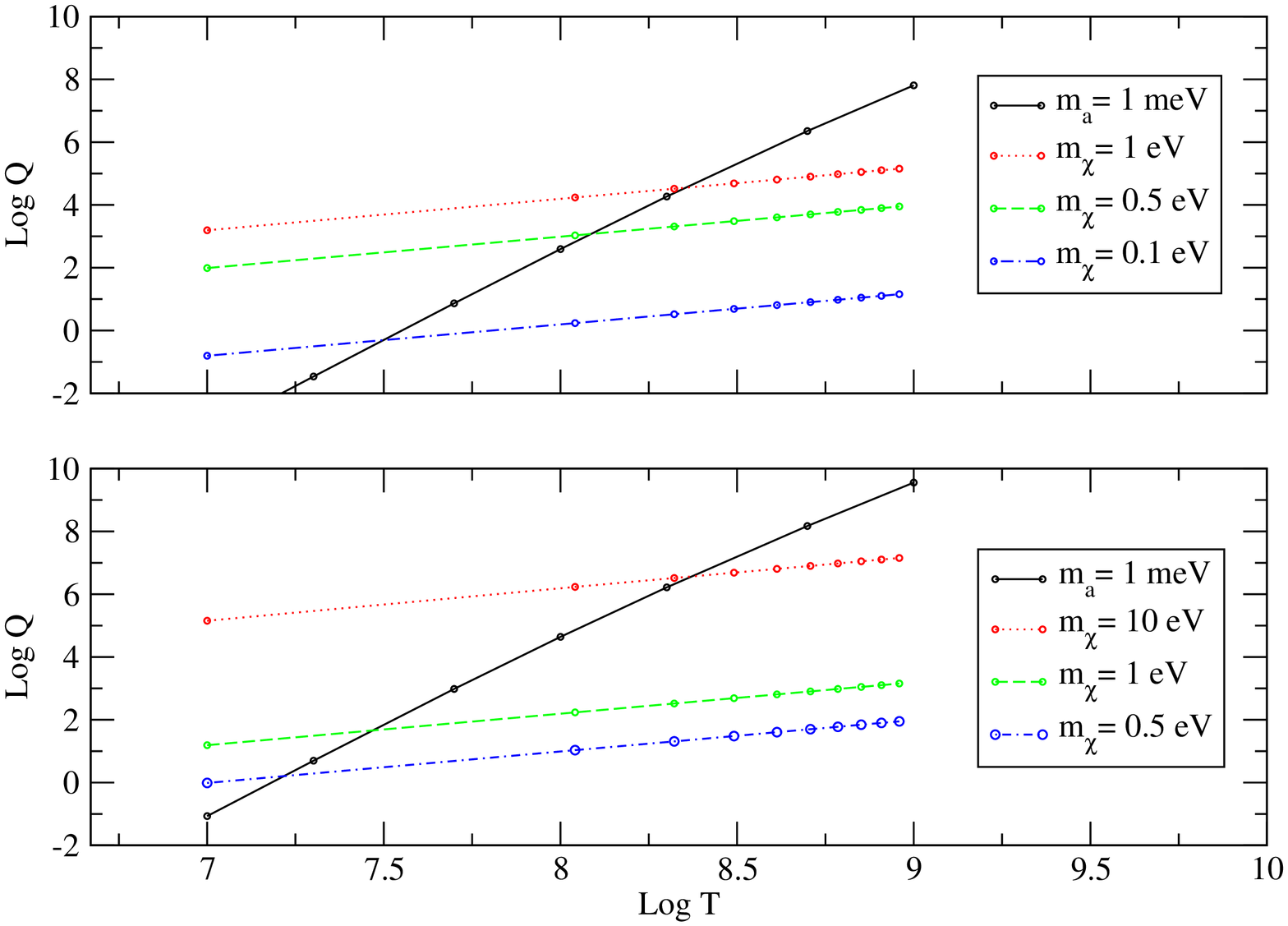}
  \caption{Comparison of the energy loss for an axion of mass $m_a$=1 meV and a 
fermion singlet $\chi$\,: $\rho\sim 10^4$ g/cm$^3$ (upper),
 $\rho\sim 10^6$ g/cm$^3$ (lower). Q in ergs/g.sec and T in K. }
\label{el2}
\end{figure}

The input parameters for the energy loss $Q$ are set for
two initial densities as $\rho \sim 10^4$ g$\cdot$cm$^{-3}$ (typical red giant) and $\rho \sim
10^6$ g$\cdot$cm$^{-3}$ (typical white dwarf). The millicharge is fixed at  
 $\xi = 10^{-3}$ and $\alpha = 1/137$. The two axion
reference masses $m_a$ are 1 eV and 1 meV. For the first
case, as can be seen in figure (\ref{el}) for both densities, the
Compton scattering of dark photons off a fermion singlet
results in $Q_{\gamma'}$ comparable to the axion $Q_a$ for masses $m_\chi$
of 1 eV, 10 eV and 1 keV. In particular, for the typical
white dwarf temperature zone ($T \sim 10^7$ K), the energy
loss from axion Compton scattering is comparable to a
dark photon scattered off a very light fermion singlet of
$m_\chi = 10$ eV. In the second case, the extremely light axion produces, for both densities, lower $Q_a$ curves, which
implies that for a comparable $Q_{\gamma'}$ , as seen in figure (\ref{el2}),
the fermion singlet masses $m_\chi$ must be smaller. Again,
for the typical white dwarf temperature zone the dark
fermion must have a mass of 0.5 eV.

\begin{figure}[t]
\centering
\includegraphics[angle=0,width=1\columnwidth ]{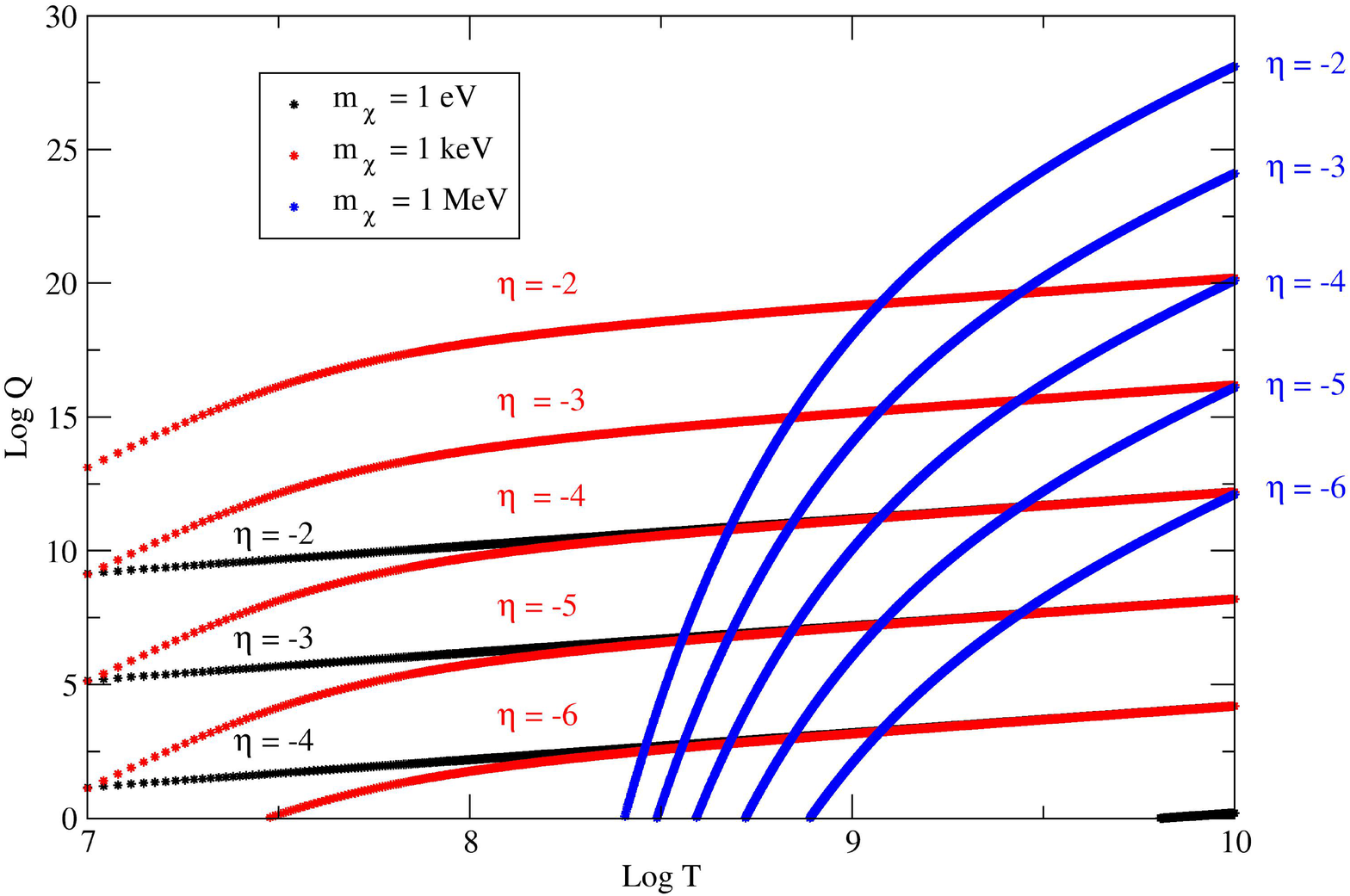}
  \caption{
Energy loss for  fermion singlets $\chi$ of different masses and millicharges: 
 $\rho\sim 10^6$ g/cm$^3$. Q in ergs/g.sec, T in K and $\eta=\log\xi$ 
}
\label{xi-fig}
\end{figure}

As well known the window with charges of order $\xi=10^{-2}$ is opened in the models with paraphotons only, when the relic density of these particles is suppressed by annihilation \cite{art1,art2}. In our model we can  test the sensitivity of the energy loss mechanism 
to the millicharge and  fermion masses for white dwarfs.  
Figure (\ref{xi-fig}), where one  defines $\eta=\log\xi$,  shows that in a region consistent with typical white dwarf 
temperatures of $10^7$ K to $10^8$ K 
the most significant energy loss for a light mass singlet of 1 eV is for $\xi=10^{-3}$ to $\xi=10^{-2}$.
In the same region there is an overlap with a heavier fermion of 1 keV, where a new  interval of smaller millicharges
is accepted from  $\xi=10^{-5}$ to $\xi=10^{-4}$. For this mass of 1 keV the most important contributions for a possible
energy loss occurs at	edge of the charge window at $\xi=10^{-2}$. A heavier fermion of 1 MeV is out of the  
typical white dwarf  temperature range and no effect would be expected.

\begin{figure}[t]
\centering
\includegraphics[angle=-90,width=.9\columnwidth ]{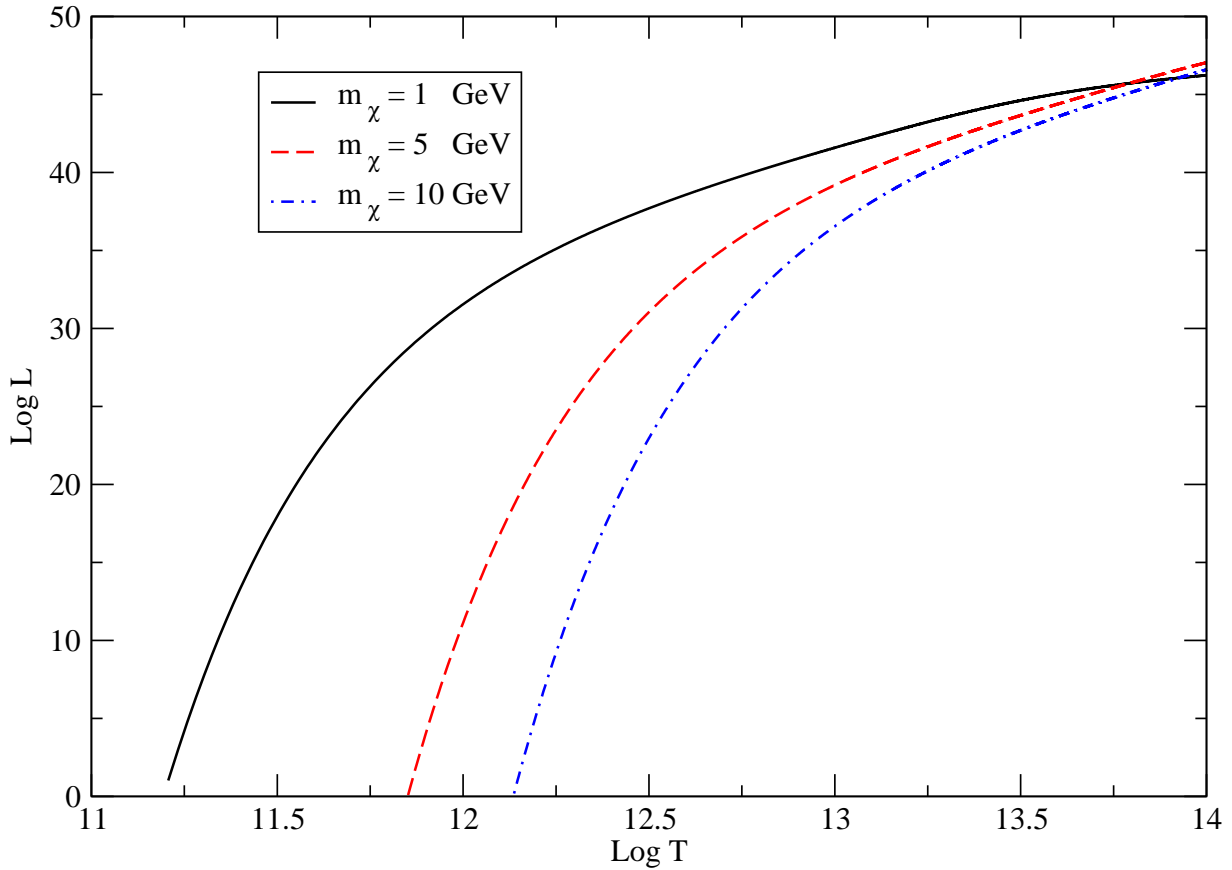}
  \caption{Average dark photon  luminosity for the fermion singlet, with masses $m_\chi=1$, 5, 10 GeV and
 $\rho\sim 10^{14}$ g/cm$^3$. L in ergs/sec and T in K. }
\label{lumi-graf}
\end{figure}

Effects of dark photons on higher density and temperature zones can also be probed in supernovas, where
typical densities are of the order $\rho \sim 10^{14}$ g$\cdot$cm$^{-3}$ and
temperatures, inside a newly formed neutron star, are
$T \sim 10^{12}$ K. The detection of neutrinos from SN 1987A
confirmed that the almost $3 \times 10^{53}$ ergs of gravitational
energy gained by the core collapse are emitted as neutrino radiation on time scales of tens of seconds, during
which the central protoneutron star (PNS) cools, deleptonizes, and contracts \cite{sn1,sn2}. In present model a very
crude luminosity estimate can be made considering a homogeneous 15 $M_\odot $ progenitor \cite{sn2,sn3}. As shown 
in Eq. (\ref{alphas})
 there is no direct coupling between the dark photon
and SM charged particles, resulting that ordinary stellar matter is transparent for dark photons. Dark photons, again, 
could be an alternative cooling mechanism
for dense matter in the neutron star regime. In the white
dwarf problem the mass range for $m_\chi$ is too restrictive
and implies in an extremely light WIMP option. For matter at neutron star densities our model admits a heavier
WIMP, as seen in figure (\ref{lumi-graf}), which is consistent with the
mass of the dark matter candidates from DAMA/LIBRA
\cite{dama-libra} and CoGeNT \cite{cdms} observations. 

A challenging issue is to find a unifying scenario in which both ``white dwarf WIMPS'' and ``supernova WIMPS" can
coexist with a consistent set of parameters. 
As can be seen in figures (\ref{el})-(\ref{lumi-graf}) there is a clear starting temperature, a {\it $Q$-threshold}, 
 which is strongly dependent on the $\chi$ mass.
If a very light $\chi$ ($m_\chi \approx 10$ keV) is present  in a supernova
it will have a $Q$-threshold at temperatures of $T\sim 10^{6}$ K, which is many orders below usual protoneutron star temperatures. To
play a significant role in the energy loss mechanism, implies that $\chi$ has a larger mass, as can be seen in figure (\ref{lumi-graf}).
A similar WIMP exclusion  occurs for a heavy $\chi$ ($m_\chi \approx 10$ GeV) in a white dwarf.  The $Q$-threshold 
will occur at temperatures of $T\sim 10^{12}$ K, again  many orders above usual white dwarf temperatures.
Therefore by this simple analysis a unifying scenario can be obtained if one  introduces  two species of dark 
fermions $\chi_1$ and $\chi_2$ in (\ref{dm1b}) with masses $m_{\chi_1}\ll m_{\chi_2}$.

\section{Conclusions}

We have studied some consequences of an extension of
the Standard Model  in a hybrid scenario where two
new $U(1)$ vector bosons were introduced. The first boson $C_\mu$ coupled to the SM by the usual minimal coupling,
producing an enlarged gauge sector in the SM, acquiring
mass by the Higgs mechanism, and a second boson $X_\mu$
mixed with $C_\mu$ via Stueckelberg coupling. After symmetry breaking, four physical bosons were present, two
massive: $Z$, $Z'$ and two photon-like $\gamma$ and $\gamma'$ (dark photon). There is a extensive literature on 
extra $U(1)$ gauge bosons that appear in the context of many models such
as $SO(10)$ or $E_6$, string and D-brane models and many
other different schemes \cite{hewett}-\cite{dm2}. Still, the cleanest signatures for a new $Z'$ boson would be mass peak 
in a resonant production in $e^+ e^-$ collision. As reported recently by the CMS Collaboration, a search for narrow
resonances was carried out in dimuon and dielectron invariant mass spectra in event samples corresponding to
an integrated luminosity of $20.6$ fb$^{-1}$ for dimuons and
$19.6$ fb$^{-1}$ for dielectrons ($\sqrt{s} = 8$ TeV) \cite{cms2}. The spectra
they found was consistent with expectations from the
SM, setting mass limits on neutral gauge bosons using
the measured dilepton spectra. In this sense, a $Z'$ with
Standard Model-like couplings has been excluded below
$2960$ GeV and the superstring-inspired below $2600$ GeV.
These results are not in contradiction with the present
model, for example Eq. (17), sets the minimum $Z'$ mass
value. In particular for $\phi = \pi/2$ (largest minimum value)
the $Z'$ mass must satisfy $m_{Z'} > 128.97$ GeV.

Direct measurements of dark photons $\gamma'$ are hopeless
due to the fact  that $\hat{\alpha}_1  = 0$ in (\ref{derivada1}). A possible scenario in which a dark photon could be important 
would be in stellar cooling. A
Compton-like diagram is present in the model converting $\gamma \leftrightarrow \gamma'$ , similar to axion models where 
$\gamma \leftrightarrow a$, and
could be used to estimate the impact in stellar cooling as
an alternative mechanism. The comparison of $Q_{\gamma'}$ and
$Q_a$ , for white dwarfs, revealed that for an extremely light
WIMP an ``overlap zone" is possible in which both mechanisms are of the same order and, in principle, could both
contribute to stellar energy loss.

There have been light-mass WIMP claims that report
an excess of low-energy events relative to expected backgrounds, the so called {\it annual modulation effect}, from
CoGeNT collaboration. This excess, if interpreted as
dark matter, implies that dark matter particles possess
a mass in the range of 5 to 15 GeV \cite{cogent} and in a recent study a mass of 7 GeV was obtained \cite{cogent2}. The
DAMA/LIBRA collaboration has presented similar results which are consistent with the CoGeNT dark matter
observations \cite{dama-libra}. In the opposite direction of these direct measurements reports, the CDMS collaboration has
recently claimed to exclude a light-WIMP interpretation
of CoGeNT and DAMA/LIBRA observations \cite{cdms}. Observations from XENON10 and XENON100 have been
used to establish a similar rejection of light-WIMP scenarios
\cite{xenon1, xenon2}. Recent results from LUX (Large Underground Xenon experiment),
have shown to be consistent with the {\it background-only
hypothesis} setting a 90\% confidence limit  
on spin-independent WIMP-nucleon elastic scattering.  A minimum upper limit on the cross 
section of 7.6 $\times 10^{-46 }$ cm$^2$ at a WIMP mass
of 33 GeV is presented \cite{lux}. Similar to other  xenon based experiments, the LUX data is in strong disagreement 
with low-mass WIMP signal
interpretations of the results from several recent direct detection experiments.

The mass of our  ``white dwarf WIMP" is far below
the measured values in these experiments. In fact, for
the energy loss mechanism, formerly described, to take
place, the dark matter fermion singlet must be lighter
than an electron ($m_\chi \ll m_e$ ). In a typical Bhabha scattering, for example, there should be sufficient energy in
the annihilation diagram to produce, in the final state,
$e^+ e^{-}$ and $\bar{\chi}\chi$.
By this argument these extremely light
dark fermions should be abundantly produced. Now, a
simple comparison of the $e^+ e^- \rightarrow \bar{\chi}\chi$  cross section with
the $e^+ e^- \rightarrow e^+ e^-$ annihilation cross section in the limit $\sqrt{s} \gg 2 m_e$,
\beann
\frac{\sigma_{e^+ e^- \rightarrow \bar{\chi}\chi}}{\sigma^{ann}_{e^+ e^- \rightarrow e^+ e^-}} \simeq \xi^2,
\eeann
showing a strong suppression of $\chi$ production. 

Alternatively, we have shown that the ``supernova WIMP" has a mass range of a few GeV, consistent 
with the claimed light-WIMP candidates and resulting in an energy-loss of the order of neutrino cooling. This
could be a promising path for future calculations. A unifying scenario where both   neutron star constraints and white dwarf
constraints are consistent implies in introducing two species of fermions $\chi_1$ and $\chi_2$ in $\call_{\chi}$ 
 with masses $m_{\chi_1}\ll m_{\chi_2}$.

\section*{Acknowledgments}
 This research was supported by 
Conselho Nacional de Desenvolvimento Cient\'{\i}fico e Tecnol\'ogico (CNPq),
Universidade Federal do Rio Grande do Sul (UFRGS).

\end{document}